# The Designability of Protein Structures:
# A Lattice-Model Study using the Miyazawa-Jernigan Matrix

Hao Li[*], Chao Tang[†], Ned S. Wingreen

*NEC Research Institute, Princeton, New Jersey*

[*]Present address: Department of Biochemistry and Biophysics, University of California at San Francisco, San Francisco, CA 94143

[†]Correspondence to: Chao Tang, NEC Research Institute, 4 Independence Way, Princeton, NJ 08540. E-mail: tang@research.nj.nec.com

# Abstract

We study the designability of all compact 3×3×3 and 6×6 lattice-protein structures using the Miyazawa-Jernigan (MJ) matrix. The designability of a structure is the number of sequences that design the structure, i.e. sequences that have that structure as their unique lowest-energy state. Previous studies of hydrophobic-polar (HP) models showed a wide distribution of structure designabilities. Recently, questions were raised concerning the use of a 2-letter (HP) code in such studies. Here we calculate designabilities using all 20 amino acids, with empirically determined interaction potentials (MJ matrix), and compare with HP model results. We find good qualitative agreement between the two models. In particular, highly designable structures in the HP model are also highly designable in the MJ model--and vice versa--with the associated sequences having enhanced thermodynamic stability.

# Introduction

The sequences and structures of natural proteins form very special classes among all possible sequences and structures. A natural protein sequence has, as its native state, a unique global minimum of free energy which is well separated in energy from other misfolded states[1]--a property not typically shared by random sequences of amino acids. Protein structures generally possess striking geometrical regularities,[2,3] characterized by preferred secondary structures and motifs[4] and often by tertiary symmetries. It has been noted that a large number of proteins are accounted for by a small number of folds,[5,6] or superfolds.[7] Several authors have proposed possible physical mechanisms behind nature's selection of protein folds. Finkelstein and coworkers argued that certain motifs are easier to stabilize and thus more common either because they have lower structural (e.g. bending) energies or because they have unusual energy spectra over random sequences.[8,9,10] Yue and Dill observed in a lattice hydrophobic-polar (HP) model that protein-like folds are associated with sequences that have a minimal number of degenerate lowest-energy states.[11] Govindarajan and Goldstein suggested that the evolutionary pressure on protein structures is to fold fast. They studied the "foldability" of structures in a lattice model and found that the foldability, optimized over sequences, varies from structure to structure.[12,13] They further argued that structures with larger optimal foldability should tolerate more sequences and be more robust to mutations.

More recently, this issue has been investigated from the perspective of "designability".[14,15,16,17] The designability of a structure is defined as the number of sequences that can design the structure--that is, sequences that possess the structure as their unique lowest-energy state. Li et al. studied the designability of all compact structures in HP lattice models of sizes 3×3×3 and 6×6.[14] They found that structures differ drastically in their designabilities and that a small number of structures emerge with designabilities much larger than the average. The sequences associated with these highly designable structures are also thermodynamically more stable[14,15] and fold much faster than typical sequences.[16] Furthermore, these structures possess regular secondary structures and motifs, and, in some cases, global symmetries.[18] Studies of designability for a larger lattice model (4×3×3)[19] and for an off-lattice model[20] yielded similar overall results.

However, most studies of designability have been based on HP-type models. It is a legitimate concern to ask how the designability of structures depends on interaction potentials and on the alphabet size (the number of different kinds of amino acids in the model).[21,22,23] In a recent lattice-model study, it was concluded that the designability of a structure depends sensitively on the size of the alphabet in the model--in particular, structures that were highly designable for a two-letter alphabet were found not especially designable with a many-letter alphabet.[22] In this paper, we study the designability of all compact structures in two lattice models using all 20 amino-acid types, with interactions given by the Miyazawa-Jernigan matrix.[24] We compare the results with those of Ref. 14 which were obtained using only two types (H and P) of amino acids. We find that the designability of a structure is *not* sensitive to the alphabet size when a realistic interaction potential (MJ matrix) is employed.

## Models and Methods

We study a lattice-protein model in both two-dimensions (2D) and three-dimensions (3D). In the 2D case, the model is a self-avoiding chain of $N=36$ residues on a square lattice. We consider only the maximally compact structures, i.e. structures contained in a 6×6 square. We also study a 2D system with a chain length $N=30$ (6×5). In the 3D case, the chain has a length of $N=27$ and folds into a maximally compact configuration of 3×3×3. The study of 3D structures is computationally limited to short chains ($N\sim30$) with corresponding surface-to-core ratios of approximately 2:1, much larger than that of typical natural proteins. 2D models can achieve a more realistic 1:1 surface-to-core ratio with manageable chain lengths, at the risk of introducing other unphysical effects due to the dimensional reduction. Thus, it is more convincing to draw conclusions based on a combined study of 2D and 3D models.

In the model, a sequence of length $N$ is specified by the residue type $\mu_i$, ($i=1, 2, \ldots, N$) along the chain, where $\mu$ is one of the 20 natural amino acids. A structure is specified by the position $\mathbf{r}_i$, ($i=1, 2,\ldots, N$) of each residue along the chain. The energy for a sequence folded into a structure is taken to be the sum of the contact energies, that is

$$E = \sum_{i<j} e_{m_i m_j} \Delta(\mathbf{r}_i - \mathbf{r}_j), \qquad (1)$$

where $e_{m_i m_j}$ is the contact energy between residue types $\mu_i$ and $\mu_j$, and $\Delta(\mathbf{r}_i - \mathbf{r}_j)=1$ if $\mathbf{r}_i$ and $\mathbf{r}_j$ are adjoining lattice sites with $i$ and $j$ not adjacent along the chain, and $\Delta(\mathbf{r}_i - \mathbf{r}_j)=0$ otherwise. The contact energies $e_{\mu\nu}$ are taken from the Miyazawa-Jernigan (MJ) matrix.[24] Note that there are several different MJ matrices in Ref. 24. We use matrix $e_{ij}$ (the upper half and diagonal of Table V in Ref. 24). This matrix contains all the contributions to the interaction energy including, in particular, the hydrophobic or solvation contribution. The hydrophobic contribution, although nonspecific, is residue dependent and is the dominant contribution to the MJ matrix $e_{ij}$.[25] For other MJ matrices in Ref. 24, the hydrophobic contribution has been, to various degrees, removed. Thus, they are not appropriate for folding studies like this one (cf. Discussion and Conclusion). We have also used updated versions of the same MJ matrix.[26,27] The results are similar.

In each case, we enumerate all maximally compact self-avoiding structures. We then randomly select a large number of sequences. For each of these sequences, we evaluate its energy on all the structures using Eq. (1). If the sequence has a unique lowest-energy state, or ground state (the criterion of being unique will be defined below), we say the sequence can design the structure and the following quantities are recorded: the ground-state structure, the ground-state energy $E_0$, the second lowest energy $E_1$, the depth of the ground state

$$\Delta = <E>' - E_0, \qquad (2)$$

and the variance of the energy spectrum

$$\Gamma^2 = <E^2>' - <E>'^2, \qquad (3)$$

where $<\bullet>'$ denotes averaging over all compact structures other than the ground state. The quantities $\Delta$, $Z=\Delta/\Gamma$,[28,29] and $\Delta_{10}=E_1-E_0$[30,31] have been widely used to characterize how protein-like a sequence is, because of their correlations with the folding rate.[32,33,16] The ground state of a sequence is said to be unique if for the sequence there are no other structures with energy lower than $E_0+g_c$, where the gap cutoff $g_c$ is a parameter. We have used $g_c$=0, 0.4, 0.8 (in the unit of RT at room temperature) in our calculations. After the calculation is completed with all randomly selected sequences, we measure the designability of a structure, $N_S$ by the number of sequences that design the structure.

We compare our results with those of Ref. 14, which were obtained using an HP model. The parameters used in Ref. 14 for Eq. (1) are: $e_{HH}=-2.3$, $e_{HP}=-1$, and $e_{PP}=0$, which were derived from and can be viewed as the two-letter simplification of the MJ matrix $e_{ij}$.[25,14]

## Results

First, we present results for the 2D 6×6 system. There are 28,728 maximally compact structures unrelated by symmetries of rotation, reflection, or reverse-labeling. In the calculation with the MJ matrix, we used up to 9,095,000 randomly selected sequences of 20 amino acids. We found that 96.74%, 42.46%, and 17.79% of sequences had a unique ground state when the gap cutoff $g_c$ was set to 0, 0.4, and 0.8, respectively. In Figure 1(a)-(c), we plot the histogram of the designability $N_S$, i.e. the number of structures with a given $N_S$ versus $N_S$. As in the case of the HP model (shown in Figure 1(d)), the distribution of $N_S$ has a long tail, that is, there are some structures with much higher than average designability. Furthermore, for large $g_c$ (Figure 1(c)) the curve resembles that of the HP model. One measure of the thermodynamic stability of a ground state is the energy gap $\Delta_{10}$ between the ground state and the next lowest energy state. To display the correlation between thermodynamic stability and designability, we average $\Delta_{10}$ over all sequences that design a structure, and then average over all structures with a given $N_S$. This doubly averaged energy gap is plotted against designability $N_S$ in Figure 2. In both models

(MJ and HP), there is a strong positive correlation between the average gap and designability $N_S$.

In order for the designability $N_S$ to be a useful characterization of structures, it should be robust with respect to some variation in model parameters. We found a very good correlation between the $N_S$'s of a given structure obtained with various gap cutoffs $g_c$ (Figure 3(a)), and obtained with the HP model (Figure 3(b)). In particular, highly designable structures in the HP model are also highly designable in the MJ-matrix model, and vice versa. The top structure is the same for both models (Figure 4).

Do the sequences that design highly designable structures have unusual ground-state energies $E_0$, or ground-state depths $\Delta$ (Eq. (2)), or spectral widths $\Gamma$ (Eq. (3))? In Figure 5 we plot the average over sequences of $E_0$, $\Delta$, $\Gamma$, and $Z = \Delta/\Gamma$ versus $N_S$. It is clear from the figure that there are no significant correlations between $N_S$ and average $E_0$ or $\Delta$. Thus, a highly designable structure does not have a lower $E_0$ or a larger $\Delta$. On the other hand, $N_S$ correlates inversely with the average width of the spectrum $\Gamma$ and therefore correlates positively with the $Z$ score. However, the scatter of data for structures of given $N_S$ is so large that small $\Gamma$ does not necessarily imply large $N_S$. A small spectral width $\Gamma$ is a necessary but not a sufficient condition for a structure to be highly designable.

To see how the distribution of sequences among the structures changes with the length of the chain, we also studied a 6×5 system. In this case, there are 6802 maximally compact structures unrelated by symmetries. We used 5,200,000 randomly selected sequences. The percentage of sequences that had a unique ground state was 96.86%, 43.96%, and 19.11%, for $g_c$=0, 0.4, and 0.8, respectively. These percentages are slightly larger than those in the 6×6 system, indicating a slight decrease in the probability of a unique ground state with increasing chain length. In HP models, the histograms of designabilities for different system sizes were found to be identical after rescaling. To test for this property in the MJ model, we let $N(N_S,L)$ be the number of structures with designability $N_S$ in the system of chain length $L$. The dependence of $N(N_S,L)$ on $L$ may be "scaled-out", and $N(N_S,L)$ may be reduced to a "universal" form. We make the scaling *ansatz* (guess)

$$N(N_S,L) = \frac{N_c}{<N_S>} f(\frac{N_S}{<N_S>}), \qquad (4)$$

where $N_c$ is the total number of structures and $<N_S>$ the average designability for chain length $L$. If Eq. (4) holds, then the "universal" function f(x) should be independent of $L$. In Figure 6, we plot f=$N<N_S>/N_c$ versus x=$N_S/<N_S>$ for systems of 6×6 and 6×5. The two curves match very well, supporting the scaling *ansatz* (4).

We now turn our attention to the 3D 3×3×3 system. There are 51,704 compact structures unrelated by symmetries. 13,550,000 randomly selected sequences of 20 amino acids were used in the calculation. With the gap cutoff $g_c$=0, 0.4, and 0.8, the percentage of the sequences that had a unique ground state was, respectively, 96.67%, 30.20%, and 8.26%. In the HP model this percentage is 4.75%.[14] Histograms of the designability $N_S$, along with the histogram for the HP model, are plotted in Figure 7. Similar to the 2D case, there is a long tail to the distribution and the histogram for $g_c$ =0.8 resembles that of the HP model. In Figure 8, we show the average gap $<\Delta_{10}>$ versus $N_S$. Again, the sequences that design structures with larger $N_S$ have larger gaps, on average. The designability of structures is rather robust characterization--we observe good correlations between $N_S$'s obtained with different $g_c$'s and between the MJ and HP models (Figure 9). The most designable structure in the MJ model (shown in Figure 10(a)) is not the same as in the HP model (shown in Figure 10(c)), though they share some common geometrical features, e.g. many anti-parallel long lines. In Figure 11 we plot the quantities $E_0$, $\Delta$, $\Gamma$, and $Z= \Delta/\Gamma$ for the 3D 3×3×3 system versus $N_S$. Similar to the 2D case, there is little dependence of $E_0$ and $\Delta$ on $N_S$. On average, there is an inverse correlation between $\Gamma$ and $N_S$, and therefore a positive correlation between $Z$ and $N_S$. The scatter of the data is large.

    Finally, we consider the set of sequences that design the top 3×3×3 structure in the MJ model. Out of 13,550,000 randomly selected sequences, 1721 of them design the top structure, namely they have the top structure as their unique ground state. In Figure 12, we plot the average hydrophobicity of the residue as a function of the

chain index *i*, averaged over all the 1721 sequences that design the top structure. It is clear that there is a strong correlation between the average hydrophobicity of the residues and the exposure to water of the site--the more buried the site, the more hydrophobic the residue, on average. In Figure 13 we plot several quantities versus the ground state energy $E_0$ for the sequences that design the top structure. We see that there is no correlation between the gap $\Delta_{10}$ and the ground state $E_0$, whereas $E_0$ is inversely correlated with both $\Delta$ and $\Gamma$. However, no obvious correlation is seen between $E_0$ and $Z=\Delta/\Gamma$, as if the effect of a lower $E_0$ is just to uniformly pull down the energy spectrum, enlarging $\Delta$ and $\Gamma$ by the same factor. Similar statistical behaviors are found for all sequences.

## Discussion and Conclusion

The above results show no sensitive dependence of designability on the alphabet size. Recently, Buchler and Goldstein studied the designability for structures on a 5×5 lattice, using various alphabet sizes for the sequence.[22,23] They obtained very poor or no correlation between the designability $N_S$ obtained with our HP parameters and with an MJ matrix. The reason for this discrepancy is that they used a different MJ matrix than the one we used to derive our HP parameters. Note that there are several matrices in Miyazawa and Jernigan's original papers.[24,26] The one we have used for this study, and for deriving our HP parameters, is the matrix $e_{ij}$, which is the upper half of Table V in Ref. 24 or the upper half of Table 3 in Ref. 26.

This is the matrix containing all interactions including the hydrophobic interaction. We have analyzed this matrix via eigenvalue decomposition,[25] and found that the matrix can be well approximated by the following form

$$e_{ij} \approx \tilde{e}_{ij} = h_i + h_j + c(i, j). \tag{5}$$

The additive term $h_i + h_j$ originates from the hydrophobic interaction and it dominates the potential (5).[25] The "two-body" term $c(i, j)$ is small compared to the additive term and represents the tendency of similar amino acids to segregate.[25] The choice of $e_{HH}$=-2.3, $e_{HP}$=-1, and $e_{PP}$=0 in our HP study can be viewed as the result of a hydrophobic part $h_H = -1$ plus a small two-body part $c(H,H) = -0.3$, with $h_P = 0$ and $c(H,P)=c(P,P) = 0$. The current study shows that there is no qualitative difference between our two-letter HP model and the full MJ matrix as far as designability is concerned. Thus, the designability of structures has no significant dependence on the alphabet size, as long as the potential is dominated by the hydrophobic or solvation force.[23] However, the outcome can be very different for qualitatively different amino-acid interaction potentials. For example, in the MJ matrix that Buchler and Goldstein used in their calculation (Table VI of Ref. 24) the hydrophobic force has been removed. It is a very different potential, dominated by the pairing term $c(i,j)$ of Eq. (5). Its set of highly designable structures is very different from that of the full MJ matrix, and is similar to that obtained for a *random* pairing potential.[23] Several authors have investigated the effect of the two-body pairing term in Eq. (5) on designability.[35,36,37,38,39] It is would be revealing to study how the designability of structures changes as the potential is changed from solvation-like to random-pairing-like.[40] It is not yet clear what role the alphabet size plays in the case of a random-pairing potential.[23]

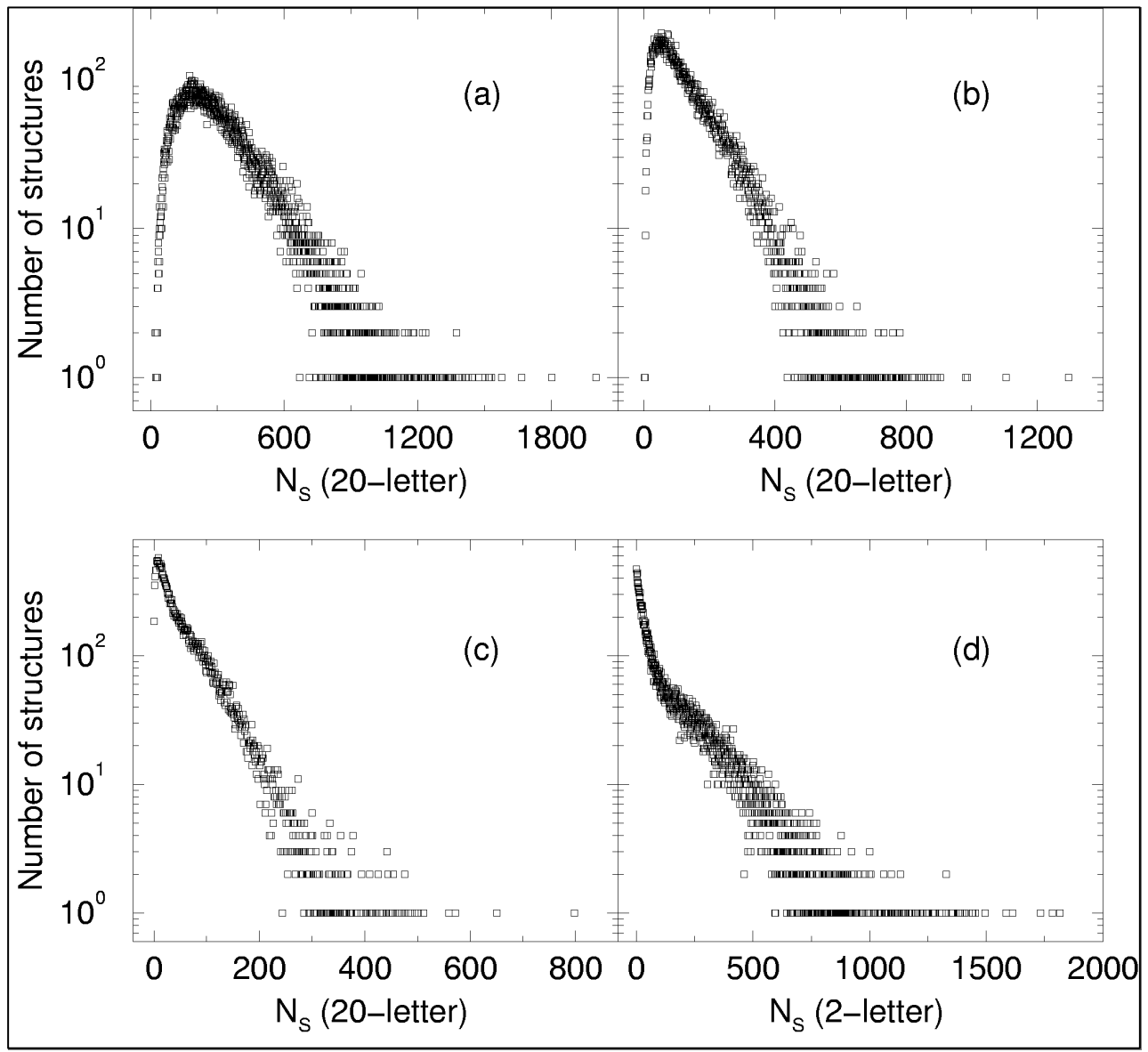

**Figure 1:** Histograms of designability $N_S$ for the 6×6 system, for the MJ matrix with gap cutoff (a) $g_c$=0.0, (b) $g_c$=0.4, (c) $g_c$=0.8, and (d) for the HP model. Results for MJ matrix were obtained by using 9,095,000 random sequences. Results for HP model were obtained by using xxxxx random sequences.

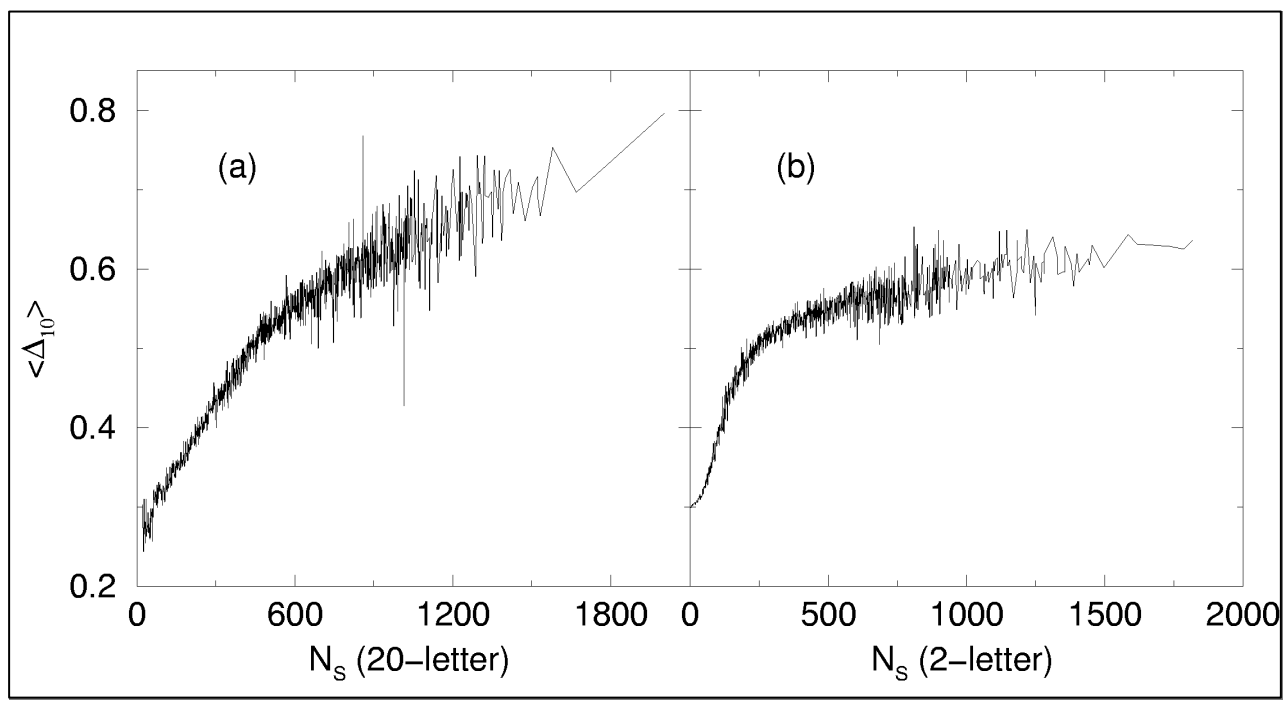

**Figure 2:** The average energy gap $\langle\Delta_{10}\rangle$ versus designability $N_S$ for the 6×6 system. (a) For the MJ matrix. (b) For the HP model.

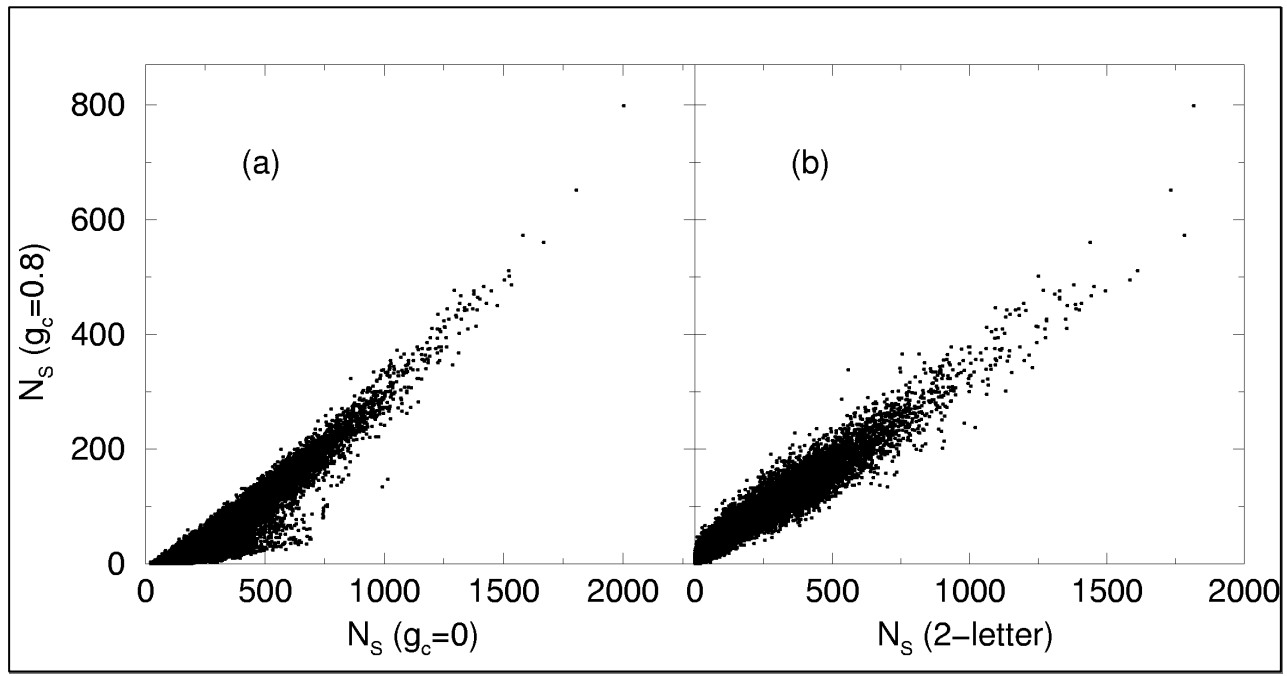

**Figure 3:** (a) Designability $N_S$ with gap cutoff $g_c=0.8$ versus $N_S$ with $g_c=0$ for the MJ model, for each 6×6 structure. (b) $N_S$ with $g_c=0.8$ for the MJ model versus $N_S$ for the HP model, for each 6×6 structure. Note that most structures are close to the origin.

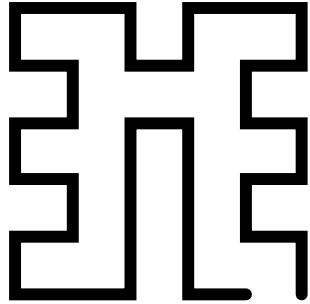

**Figure 4:**   The most designable structure in the 6×6 system.

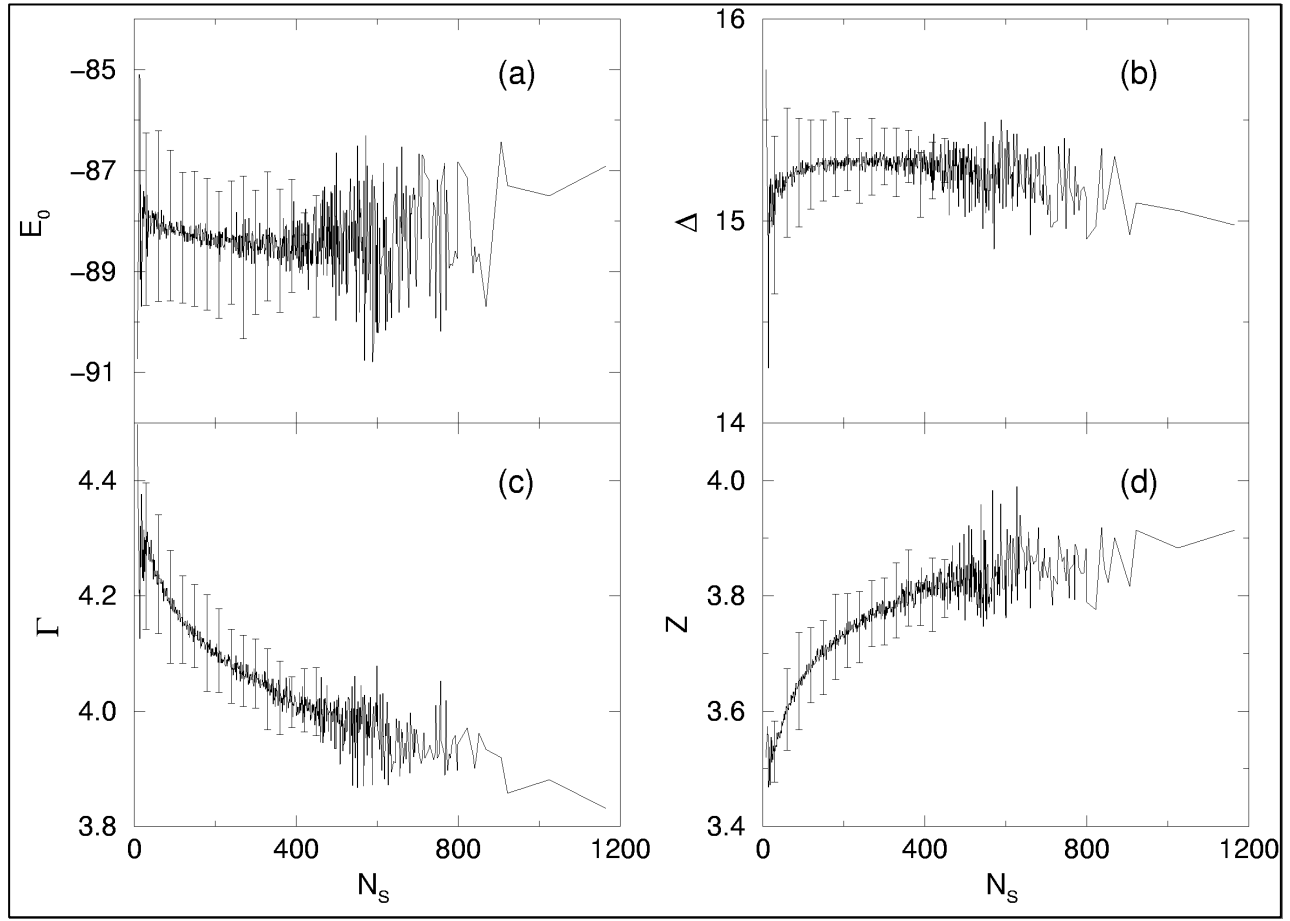

**Figure 5:** (a) Ground-state energy $E_0$, (b) depth of ground state $\Delta$, (c) width of compact spectrum $\Gamma$, and (d) $Z=\Delta/\Gamma$ versus $N_S$ for the 6×6 system with gap cutoff $g_c=0$. The solid lines are averages for given $N_S$, and the error bars indicate the variances. Data were obtained from 5,100,000 random sequences.

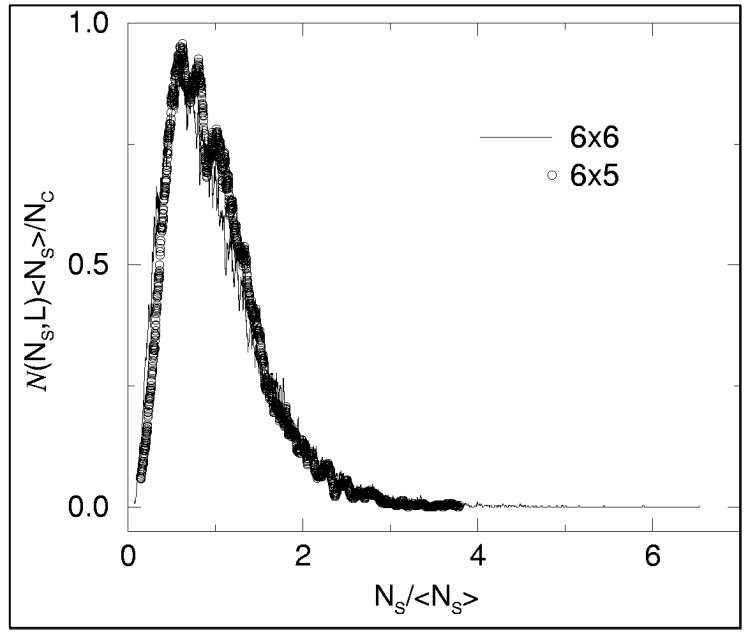

**Figure 6:** Scaled histogram versus scaled designability $N_S$ for 6×6 and 6×5 systems. For a chain of length $L$, we denote by $N(N_S,L)$ the number of structures with designability $N_S$. The average designability is $\langle N_S \rangle$ and $N_C$ is the total number of structures.

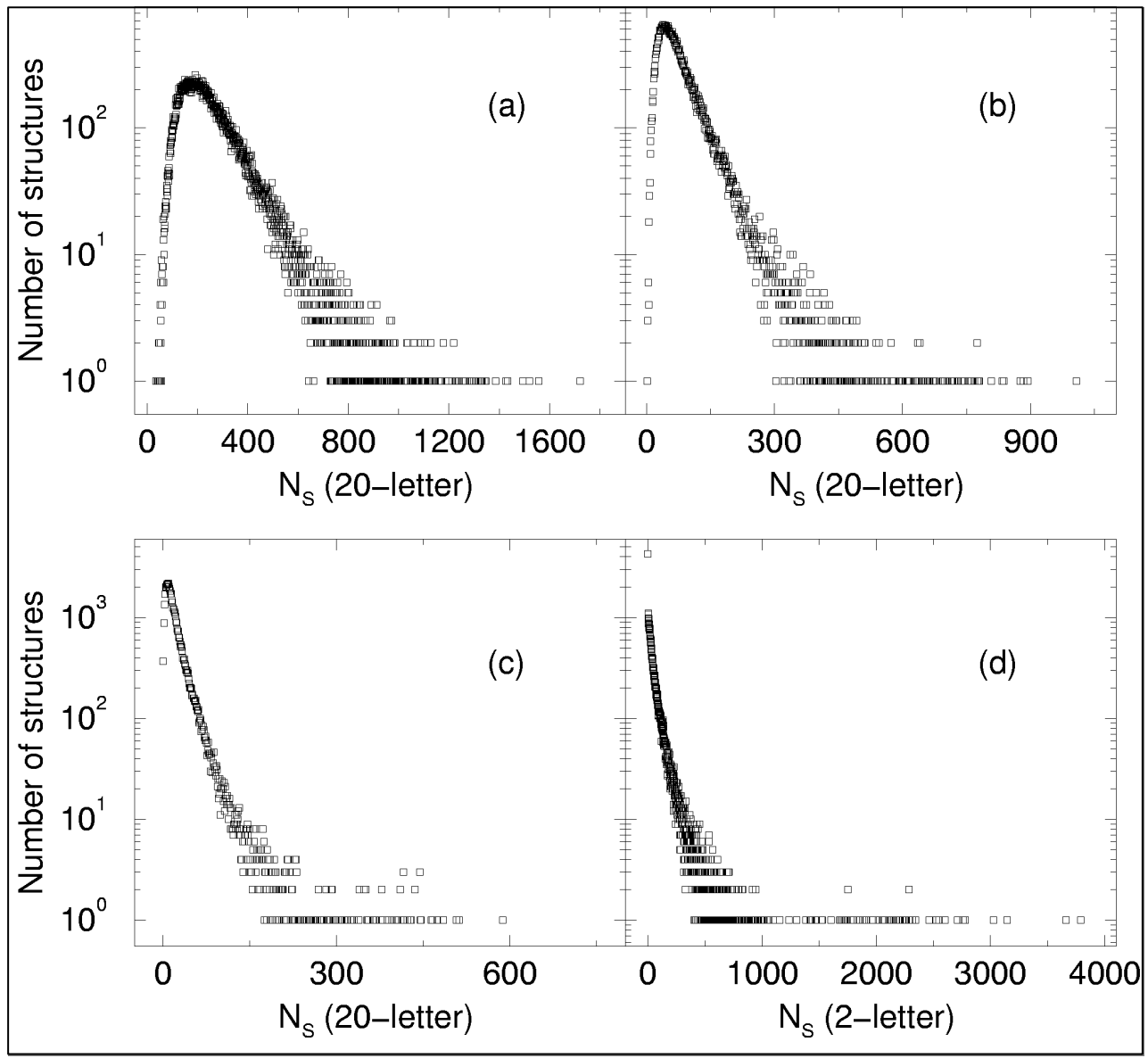

**Figure 7:** Histogram of designability $N_S$ for the 3×3×3 system, for the MJ matrix with gap cutoff (a) $g_c=0.0$, (b) $g_c=0.4$, (c) $g_c=0.8$, and (d) for the HP model. Results for MJ matrix were obtained by using 13,550,000 random sequences. Results for HP model were obtained by enumerating all $2^{27}$ sequences.

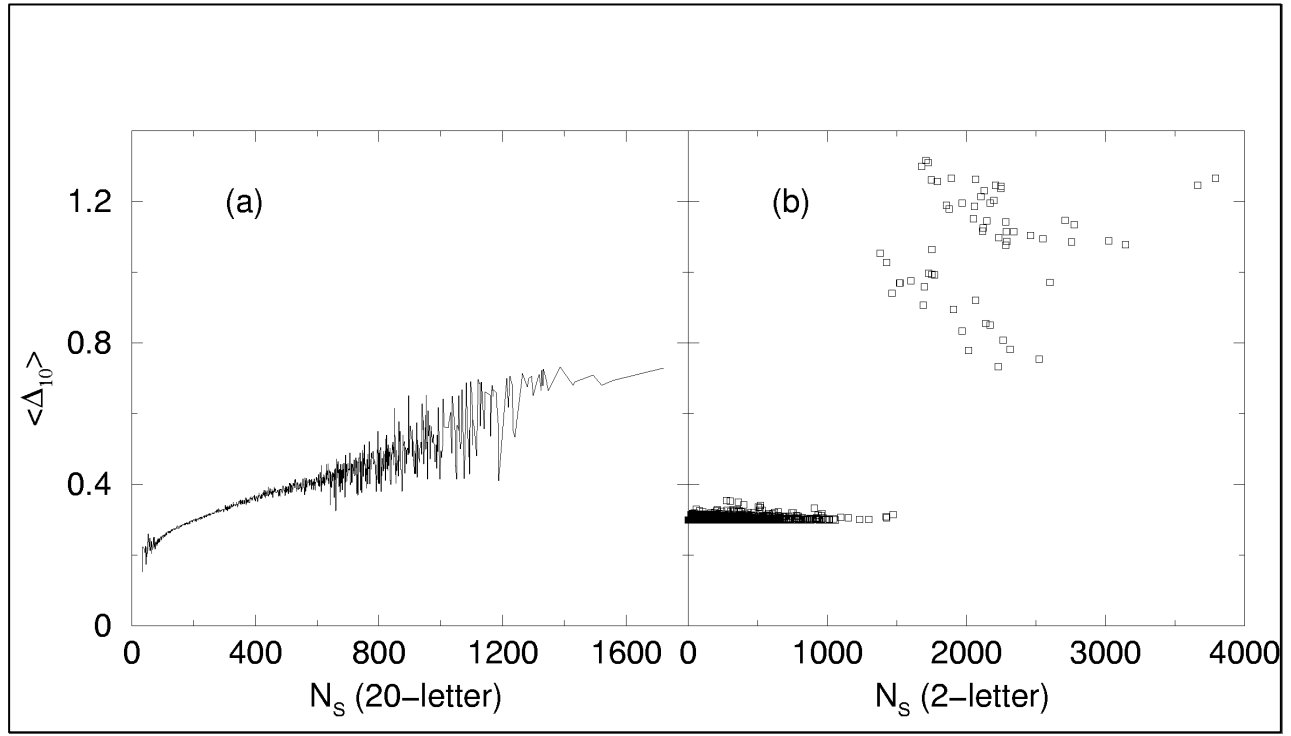

**Figure 8:** The average gap $\langle \Delta_{10} \rangle$ versus designability $N_S$ for the 3×3×3 system. (a) For the MJ matrix. (b) For the HP model.

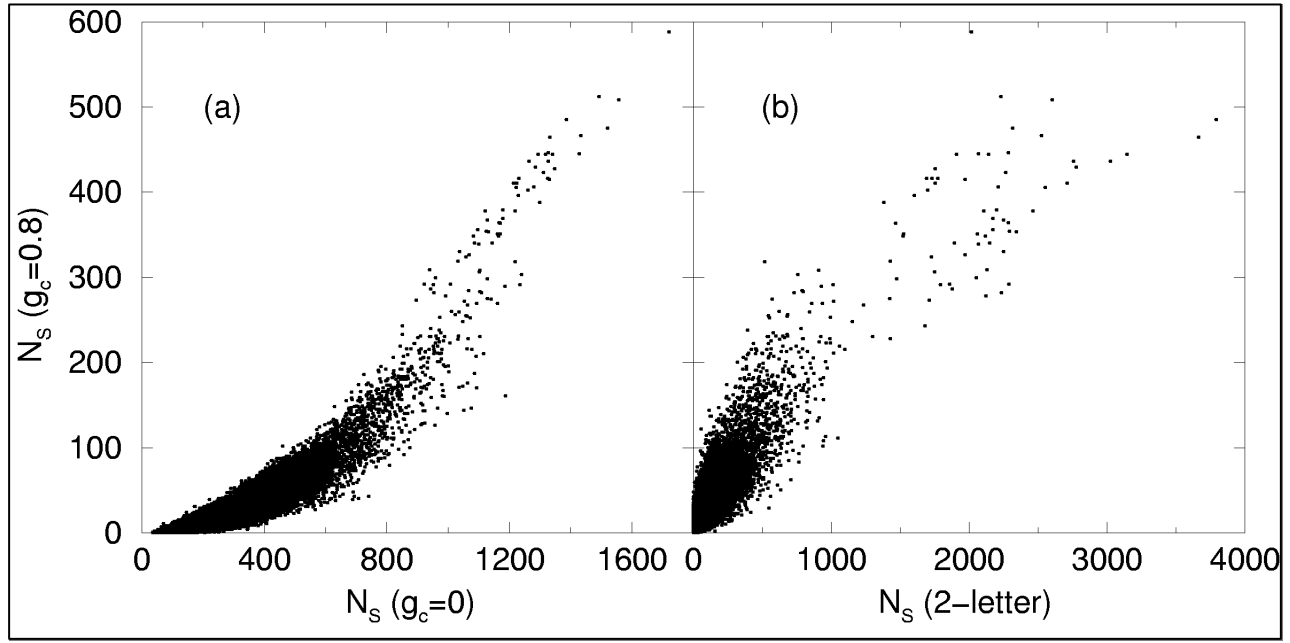

**Figure 9:** (a) Designability $N_S$ with gap cutoff $g_c=0.8$ versus $N_S$ with $g_c=0$ for the MJ matrix, for each 3×3×3 structure. (b) $N_S$ with $g_c=0.8$ for the MJ matrix versus $N_S$ for the HP model, for each 3×3×3 structure.

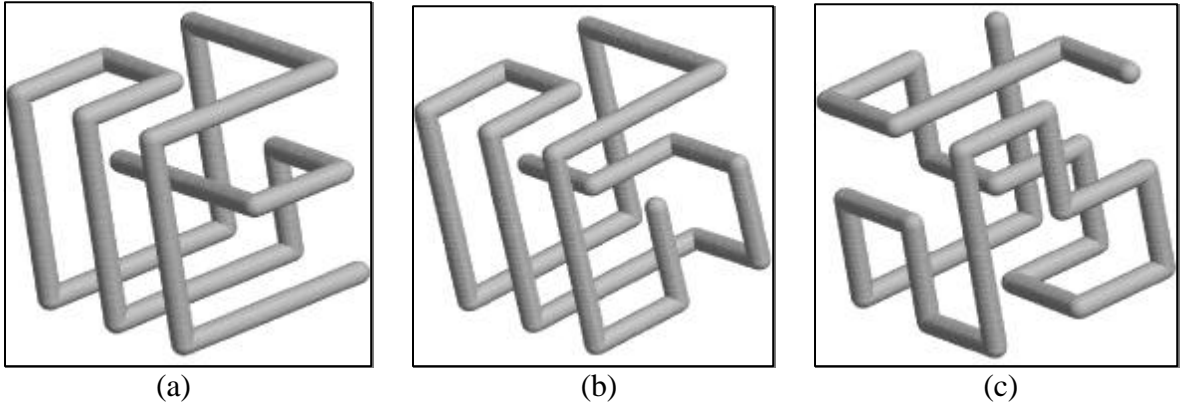

**Figure 10:** (a) The top structure for the 3×3×3 system with the MJ matrix. (b) The top structure for the HP model. (c) A structure with low $N_S$ in both the MJ and HP models. Poorly designable structures typically show less geometrical regularity than highly designable structures.

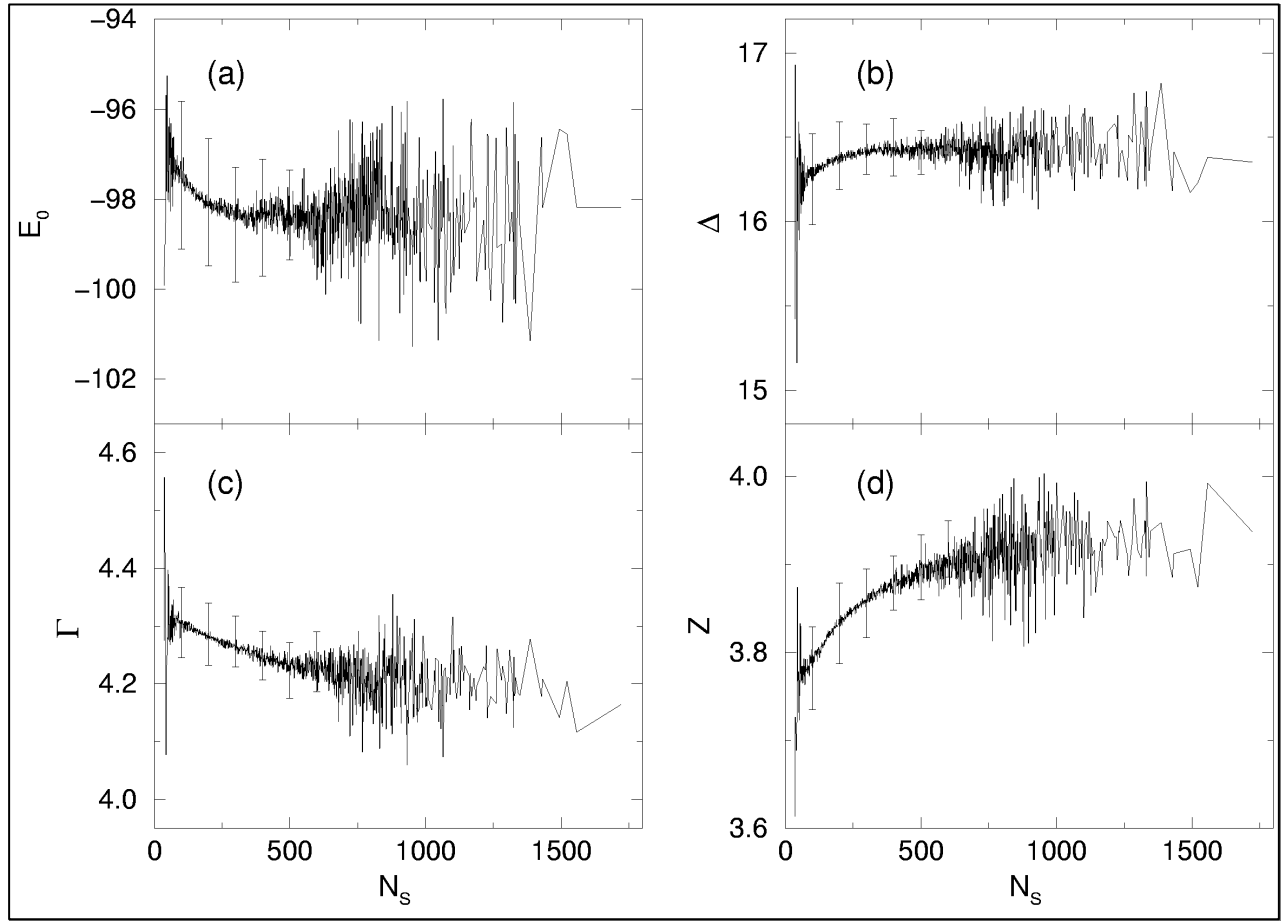

**Figure 11:** (a) Ground-state energy $E_0$, (b) ground-state depth $\Delta$, (c) width of compact spectrum $\Gamma$, and (d) $Z=\Delta/\Gamma$ versus $N_S$ for the 3×3×3 system. Solid lines are averages for given $N_S$. Error bars indicate variances.

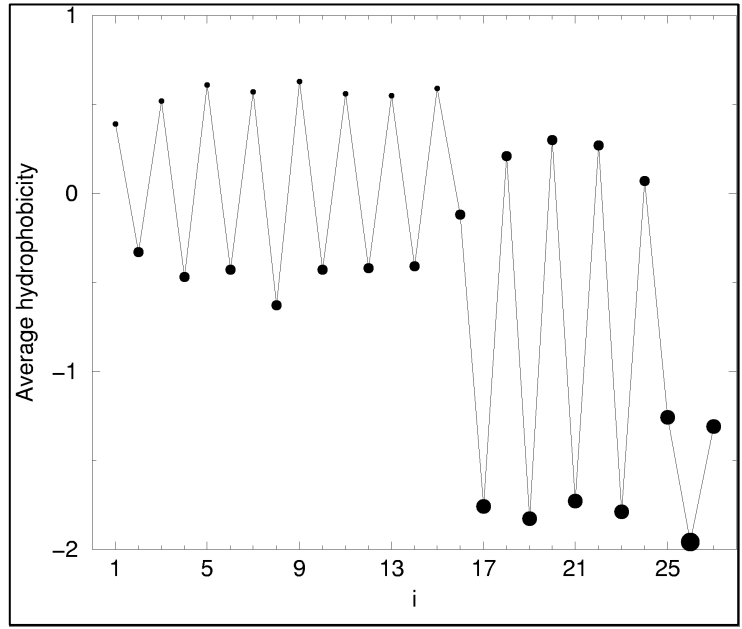

**Figure 12:** Average hydrophobicity over the sequences that design the top 3×3×3 structure in the MJ model. The size of each black dot, from small to large, represents the position of the residue: corner, edge, face, and center. The hydrophobicity scale of the 20 amino acids is taken from Creighton TE, *Proteins* (Freeman, New York, 1993).

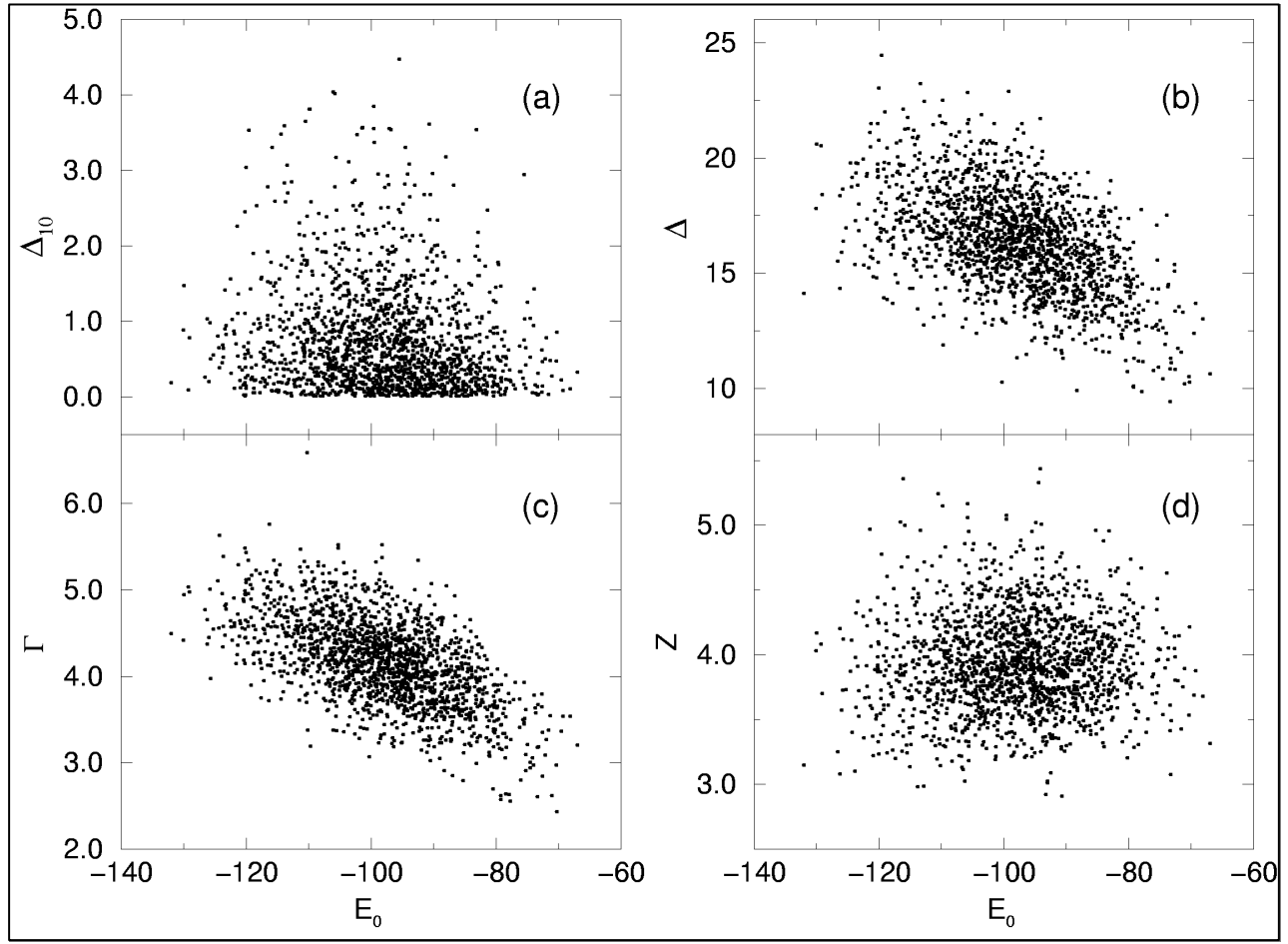

**Figure 13:** (a) Energy gap $\Delta_{10}$, (b) depth of ground state $\Delta$, (c) width of compact spectrum $\Gamma$, and (d) $Z=\Delta/\Gamma$ versus the ground-state energy $E_0$ for the sequences that design the top structure in the MJ model for the 3×3×3 system.